\documentclass[a4paper,12pt]{article}
\usepackage{epsfig}
\usepackage[dvips,usenames]{color}

\newlength{\dinwidth}
\newlength{\dinmargin}
\newcommand{\spur}[1]{\not\! #1 \,}
\begin{document}
\date{}
\title{Study of ${B}^{0}\to J/\Psi  D^{(\star)}$ and $\eta_c D^{(\star)}$ 
in Perturbative QCD}
\vskip 15mm
\author{Gad Eilam, Massimo Ladisa and Ya-Dong Yang\\
{\small Physics Department, Technion-Israel Institute of Technology,
 Haifa 32000,  Israel}\\}
\maketitle


\begin{abstract}

Motivated by recent interest in soft $J/\Psi$ production in $B$ decays, 
we investigate ${B}^0 \to J/\Psi D^{(\star)}$ and $\eta_c D^{(\star)}$
decays in perturbative QCD. We find that, within that framework,
these decays are calculable since  
the heavy $c\bar{c}$ pair  in the final states is created by a hard gluon.
The branching ratios are estimated to be around 
$10^{-7}\sim 10^{-8}$, too small to be consistent with
the data, suggesting that other mechanism(s) contribute to the observed
excess of soft $J/\Psi$ in ${B}^0 \to J/\Psi + X$
decays. The possibility of the production of a hybrid 
$s{\bar d}g$ meson with mass about $2$ GeV is briefly
entertained.

{\bf PACS Numbers: 12.38Bx, 12.39Jh}

{\bf Keywords: $B$-meson, charmonium, $D$ meson,
$D^\star$ meson, hybrid meson}
\end{abstract}
\newpage

With the advent of the BaBar and Belle $B$ factories, many $B$ 
decay modes could be studied in 
detail. The rich phenomena of $B$ decays will provide testing grounds 
for theories of weak interactions and hadrons. It is interesting to note
that measurements of the inclusive $B\to J/\Psi X $ spectrum by CLEO 
\cite{CLEO}
and recently by Belle
\cite{Belle}, indicate a 
hump for low $J/\Psi$ momentum, which kinematically corresponds to $J/\Psi$
recoiling against a partner as heavy as $\sim 2~{\rm GeV}$. 
Brodsky and Navarra \cite{brod}
suggest that the $J/\Psi$ hump may be due to the decay ${B}^{0}\to 
J/\Psi \Lambda\bar{p}$ with possible formation of $\Lambda-\bar{p}$ bound 
state (an exotic strange baryonium).   

From another view point, Chang and Hou
\cite{hou} 
proposed as an explanation the existence of 
intrinsic charm in the $B$ meson which decays as
${B}^{0}(\bar{d}bc\bar{c})\to J/\Psi D^{(*)}$ (and similarly
for $\eta_c$ instead of $J/\Psi$). 
Thus the intrinsic
charm pair transforms into a $c {\bar c}$ final state while the $b$ decays. 
It is argued 
that a rate of $\sim 10^{-4}$ may be possible in this way if the intrinsic 
charm content of $B$ is not much less than $1\%$. 

We raise here another possibility: The $B$ may decay into a 
charmonium plus a hybrid, $B^0 \to J/\Psi H$, where $H$ is a hybrid
$s {\bar d}g$ \cite{Davies+Tye}
with $M_H \approx 2~{\rm GeV}$ \cite{hybrid}. Two diagrams that contribute to
such a process are depicted in Fig. 1. Note that the gluons exchanged
in Fig. 1 are soft while those in Fig. 2 ({\it i.e.} for the
conventional $B^0 \to J/\Psi D^{(*)}$, see below) 
are hard, thus enhancing 
the hybrid option as compared to the conventional approach for $B^0 \to J/\Psi
 D^{(*)}$.
In addition, as shown below, each Feynman diagram in 
Fig. 2 involves one fermion and one 
hard 
gluon propagator with average virtuality as large as $10~{\rm GeV}^2$. So,
 we can expect the
$B^0\to J/\Psi H$ decay rate to be $10^{3}\sim 10^{4}$ times larger than 
$B^0 \to J/\Psi D^{(*)}$,
although a reliable quantitative estimate of the decay rate is very 
difficult.

\begin{figure}[htbp] 
\begin{center}
\scalebox{1}{\epsfig{file=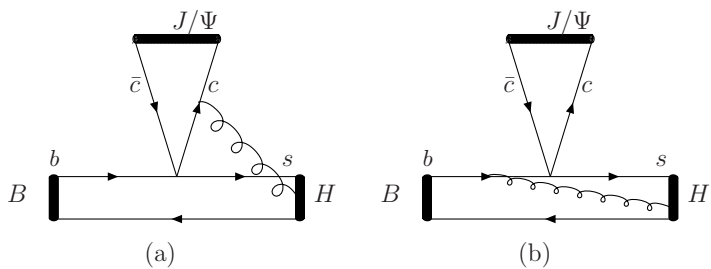}}
\end{center}
\caption{Examples of Feynman diagrams for 
the production of a hybrid $H=s {\bar d}g$
in ${B}^0 \to J/\Psi H$.}
\end{figure}

To make such ``exotic''
suggestions more reliable, one should be convinced  that the 
conventional picture of heavy mesons indeed leads to tiny numbers in 
disagreement with experiment. To our knowledge,
such study is still not available in the literature. In this article we 
investigate 
these decays within the conventional picture of heavy mesons having the 
minimal number of
quarks and using perturbative QCD 
(pQCD). The applicability of pQCD
is justified by the large virtuality of the hard 
gluon which creates a $c\bar{c}$ pair.
As known, in many applications of pQCD to B decays
\cite{appQCD}, say $B\to \pi\pi$,
the virtuality of the gluon in the hard kernel scales like 
$k^2_g \simeq-M_{B}\Lambda_{\rm{QCD}}x\simeq-2x~\rm{GeV}^2$, 
where $x$ is the momentum 
fraction carried by the light spectator quark in the final light meson. 
However in the processes discussed in this paper (see Fig. 2), 
the gluon virtuality scales
as $k^2_g >(2m_{c})^2$. Furthermore, under the common assumption of
factorization, there are no infrared divergences 
which cannot be absorbed in wave function, or large 
end point contributions.

\begin{figure}[htbp] 
\begin{center}
\scalebox{1}{\epsfig{file=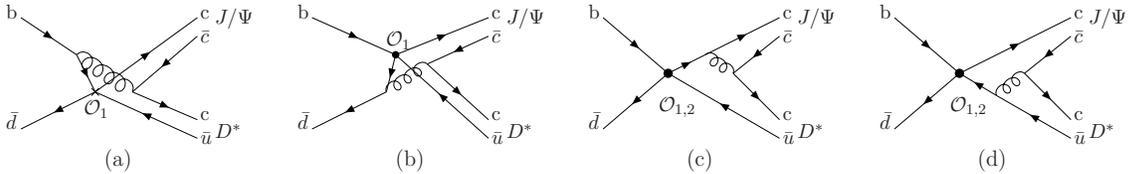}}
\end{center}
\caption{Feynman diagrams for ${B}^0 \to J/\Psi D^{(*)}$ in pQCD.} 
\end{figure} 

We begin our calculation of the decays ${B}^{0}\to J/\Psi D^{(*)}$ 
within the pQCD approach for exclusive
QCD processes \cite{BL} as depicted in Fig. 2,
by writing the weak effective 
Hamiltonian 
$H_{\rm eff}$
for the $b\to c\bar{u}d$ transitions as \cite{hamil}
\begin{equation}
\label{heff}
{\cal H}_{\rm eff}
=\frac{G_{F}}{\sqrt{2}}  V_{cb} V_{ud}^{*}
\left[
C_{1}(\mu)\bar{d}\gamma_{\mu}(1-\gamma_{5})u\, 
\bar{c}\gamma^{\mu}(1-\gamma_{5})b +  
C_{2}(\mu)\bar{c}\gamma_{\mu}(1-\gamma_{5})u\, 
\bar{d}\gamma^{\mu}(1-\gamma_{5})b  
\right],
\end{equation}
where the Wilson coefficients $C_{1,2}(\mu)$ are evaluated  to be  
$C_{1}(m_{b})=1.124$ and $C_{2}(m_{b})=-0.273$ at the scale 
$\mu=m_{b}=4.8~{\rm GeV}$ \cite{Neub}. 

 To calculate the amplitudes of the Feynman diagrams 
in Fig. 2, we take the wave 
functions for ${B}^{0}$, $J/\Psi$ and $D^*$ as follows \cite{Kuhn, appQCD}
\begin{eqnarray}
{\bf \Psi}_{B}&=& \frac{i}{4N_c} (\not P_{B} + M_{B})\gamma_5 \phi_{B}(x) 
f_{B}, \\
{\bf  \Psi}_{V}&=& -\frac{1}{4N_c} \not\varepsilon( M_{V}+\not P_V )
  \phi_{V}(x)f_{V}, ~~~(V=J/\Psi, D^{*}).
\end{eqnarray}  
Now we can write the amplitudes of Fig. 2. as 
\begin{eqnarray}
{\cal M}_a &=&-if_{B}f_{\Psi}f_{D^*}\pi\alpha_{s}\frac{1}{16}\frac{C_F}
{N^2_c}
  C_{1} 
\int dxdydz\phi_{B}(x)\phi_{\Psi}(y)\phi_{D^*}(z)\frac{1}{D_a}\frac{1}{k^2}
\nonumber\\
          &&\times {\rm Tr}[(\not{P}_{B} + M_{B})\gamma_{5}\gamma_{\mu}(1-
\gamma_5 )
           ((1-x)\not{P}_{B} - \not{k}+m_b )\gamma_{\nu}] \nonumber\\
          &&  \times {\rm Tr}[\not{\varepsilon}_{\Psi}(M_{\Psi} 
+ \not{P}_{\Psi}) 
             \gamma^{\mu}(1-\gamma_5 )\not{\varepsilon}_{D^{*}}(M_{D^*} 
              + \not{P}_{D^*})\gamma^{\nu}],\\
{\cal M}_b &=&-if_{B}f_{\Psi}f_{D^*}\pi\alpha_{s}\frac{1}{16}\frac{C_F}
{N^2_c}
  C_{1} 
\int dxdydz\phi_{B}(x)\phi_{\Psi}(y)\phi_{D^*}(z)\frac{1}{D_b}\frac{1}{k^2}
\nonumber\\
          &&\times {\rm Tr}[(\not{P}_{B} + M_{B})\gamma_{5}\gamma_{\nu} 
           (\not{k}-x\not{P}_{B})\gamma_{\mu}(1-\gamma_5 )] \nonumber\\
          &&\times {\rm Tr}[\not{\varepsilon}_{\Psi}(M_{\Psi} 
+ \not{P}_{\Psi}) 
             \gamma^{\mu}(1-\gamma_5 )\not{\varepsilon}_{D^{*}}(M_{D^*} 
              + \not{P}_{D^*})\gamma^{\nu}],\\
{\cal M}_c &=&-if_{B}f_{\Psi}f_{D^*}\pi\alpha_{s}\frac{1}{16}\left(
\frac{C_F}{N^2_c}C_{1}+\frac{C_F}{N_c}C_{2}\right)
\int dxdydz\phi_{B}(x)\phi_{\Psi}(y)\phi_{D^*}(z)\frac{1}{D_c}\frac{1}{k^2}
\nonumber\\
          &&\times {\rm Tr}[(\not{P}_{B} + M_{B})\gamma_{5}\gamma_{\mu}(1-
\gamma_5 )]
           {\rm Tr}[\not{\varepsilon}_{\Psi}(M_{\Psi} + \not{P}_{\Psi}) 
             \gamma_{\nu}(\not{P}_{\Psi} + \not{k}+m_{c})\nonumber\\
          &&   \gamma^{\mu}
             (1-\gamma_{5})
             \not{\varepsilon}_{D^{*}}(M_{D^*} 
              + \not{P}_{D^*})\gamma^{\nu}],\\
{\cal M}_d &=&if_{B}f_{\Psi}f_{D^*}\pi\alpha_{s}\frac{1}{16}\left(
\frac{C_F}{N^2_c}C_{1}+\frac{C_F}{N_c}C_{2}\right)
\int dxdydz\phi_{B}(x)\phi_{\Psi}(y)\phi_{D^*}(z)\frac{1}{D_d}\frac{1}{k^2}
\nonumber\\
          &&\times {\rm Tr}[(\not{P}_{B} + M_{B})\gamma_{5}\gamma_{\mu}(1-
\gamma_5 )]
           {\rm Tr}[\not{\varepsilon}_{\Psi}(M_{\Psi} + \not{P}_{\Psi})
              \gamma^{\mu}(1-\gamma_{5})
              \nonumber\\
          &&  ((1-z)\not{P}_{D^*} + \not{k})\ \gamma_{\nu}
             \not{\varepsilon}_{D^{*}}(M_{D^*} 
              + \not{P}_{D^*})\gamma^{\nu}],
\end{eqnarray} 
where $C_F =\frac{4}{3}$ and $N_c =3$ is the number of colors.
$D_i(i=a,b,c,d)$ and $k^2$  denote the virtuality of quark and gluon 
propagators in Fig. 2.$i$, which are given by  
\begin{eqnarray}
D_a &=&-m^2_{b}+M^2_{B}(1-x-y)(1-x-z)+
(y-z)(M^2_{\Psi}(x+y-1)-M^2_{D^*}(x+z-1)) \nonumber\\
    && -i\epsilon,\\
D_b &=&M^2_{B}(x-y)(x-z)+(y-z)(M^2_{\Psi}(y-x)+M^2_{D^*}(x-z))-i\epsilon,\\
D_c &=&-m^2_c +M^2_{\Psi}(1-z)+(M^2_{B}-M^2_{D^*}(1-z))z-i\epsilon, \\
D_d &=&\frac{1}{2}(M^2_{B}(1+2y-z)+M^2_{D^*}(1-2y+z)+
M^2_{\Psi}(2y^2 -2yz+z-1))-i\epsilon,\\
k^2 &=&M^2_{\Psi}y(y-z)+z(M^2_{B}y+M^2_{D^*}(z-y))-i\epsilon.
\end{eqnarray}
It may be instructive to evaluate
typical virtualities of the propagators involved 
in pQCD calculations. Taking $x=1-m_{b}/M_{B},\, y=1/2$ and 
$z=m_{c}/M_{D^*}$, we find 
\begin{equation}
D_a =-20.4~{\rm GeV}^{2},\,\, D_b =7.2~{\rm GeV}^{2},
\,\, D_c =20.2~{\rm GeV}^{2},\,\, D_d =16.4~{\rm GeV}^{2},\,\,
k^2 =9.9~{\rm GeV}^2 . 
\end{equation}
These values are large enough to  justify our pQCD calculation.

The amplitude for ${B}^{0}\to J/\Psi D^*$ is decomposed as 
\begin{equation}
A({B}\to J/\Psi(P_{\Psi}) D^{*}(P_{D^*}))=\varepsilon^{\mu}_{\Psi}
\varepsilon^{\nu}_{D^{*}}({\cal S}~g_{\mu\nu}+{\cal D}~P_{D^*\mu}P_{\Psi\nu} 
+i{\cal P}~
\epsilon_{\mu\nu\alpha\beta}P^{\mu}_{\Psi}P^{\nu}_{D*}),
\end{equation}
where the coefficients ${\cal S}$, ${\cal P}$ and ${\cal D}$ correspond to 
$s$, $p$ and $d$ wave amplitudes,
respectively, and can be evaluated from Eqs. 4 to 7. 
The helicity amplitudes are constructed to be

\begin{eqnarray}
H_{00}&=&\frac{1}{2M_{\Psi}M_{D^*}}\left[{\cal S}(M^2_{B}
-M^2_{\Psi}-M^2_{D^*})+2{\cal D} M^2_{B}
|{\bf p}|^2 \right], \nonumber \\
H_{\pm\pm}&=&-({\cal S}\pm {\cal P}M_{B}|{\bf p}|).
\end{eqnarray}
The branching ratio is given by 
\begin{equation}
{\rm Br}({B}^0 \to J/\Psi D^* )=\tau_{{B}^0} \frac{|{\bf p}|}{8\pi M^2_{B}}
\frac{G^2_{F}}{2}|V_{cb}|^2 \left( |H_{00}|^2 
+ |H_{++}|^2 +|H_{--}|^2  \right).
\end{equation}

Since the $b$ and $c$ quarks are heavy and 
their mass is much larger than the typical
QCD scale $\Lambda_{\rm QCD}$ for a bound state, we can expect that  
the distribution functions of 
heavy mesons will peak around the points where the heavy quarks are
near their mass shell with
variance $\Lambda_{\rm QCD}/m_Q$. As an ansatz, the distribution 
functions are taken as
\begin{equation}
\phi_{B}(x)=\delta(x-x_B ),\,\, \phi_{\Psi}(y)=\delta(y-y_{\Psi}),\,\,
\phi_{D^*}(z)=\delta(z-z_{D^*}), 
\end{equation}   
with $x_{B}=1-\frac{m_{b}}{M_B}$, $y_{\Psi}=
\frac{1}{2}$ and $z_{D^*}=\frac{m_c}
{M_{D^*}}$.

To get numerical results, we use 
 $V_{cb}=0.04,~ f_{B}=180~{\rm MeV}, ~~ 
f_{\Psi}=400~{\rm MeV},~~ f_{D^*}=230~{\rm MeV},~~ 
m_b =4.8~{\rm GeV},~ M_{{B}^0}=5.27~{\rm GeV},~ m_c =
1.4~{\rm GeV},~ M_{\Psi}=3.1~{\rm GeV},~
M_{D^*}=2~{\rm GeV},~ \alpha_s (2m_{c})=0.266$. 
We get
\begin{equation}  
{\rm Br}({B}^0 \to J/\Psi D^* )=6.46\times 10^{-8},
\end{equation}
and the longitudinal polarization fraction is 
\begin{equation}
P_{L}=\frac{\Gamma_{L}}{\Gamma}=0.398.
\end{equation}
Since the amplitudes are highly suppressed 
by  the large virtualities of the propagators as shown in Eqs. 8 to 13, 
the smallness
of ${\rm Br}({B}^0 \to J/\Psi D^{*})$ is understandable. To illustrate the 
stability of our results, we plot in Fig. 3
${\rm Br}({B}^0 \to J/\Psi D^* )$ versus
$x_B$, {\it i.e.}, the peak point  of $\phi_{B}(x)$. 

\begin{figure} 
\begin{center}
\scalebox{0.9}{\epsfig{file=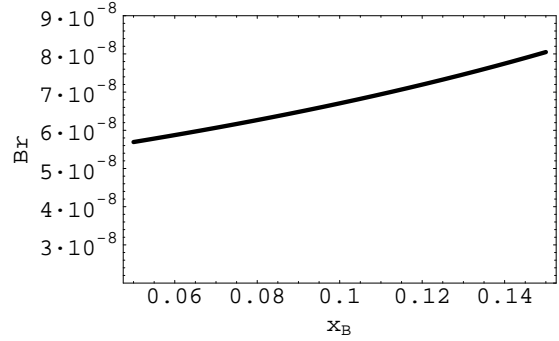}}
\caption{${\rm Br}({B}^0 \to J/\Psi D^* )$ {\it vs.} $x_{B}$, 
the peak point of $\phi_B(x)$.}
\end{center}
\end{figure} 

\ From Fig. 3, we can see that the rate is rather stable 
against changes of the parameter
$x_{B}$.  Due to relativistic effects, the distribution functions should 
have variances of ${\cal O}(\Lambda_{\rm QCD}/m_Q)$. 
To show the effects of the variances,   
we take  
\begin{eqnarray}
 \phi_{B}(x)&=&N_{B} x (1-x) \exp\left[ -\left(\frac{M_{B}}{M_{B}
-m_{b}}\right)^2
\left(x-x_{B} \right)^2 \right],\\
\phi_{D^*}(x)&=&N_{D^*} x (1-x) \exp\left[ -\left(\frac{M_{D^*}}{M_{D^*}
-m_{c}}\right)^2
\left(x-x_{D^*} \right)^2 \right],\\
\phi_{\Psi}(x)&=&N_{\Psi} x (1-x) \exp\left[ -\left(\frac{M_{\Psi}}{M_{\Psi}
-2m_{c}}\right)^2
\left(x-\frac{1}{2} \right)^2 \right],
\end{eqnarray}
where $N_{B}$, $N_{D^*}$  and $N_{\Psi}$ are  normalization constants to make
$\int dx\phi(x)=1$. To model the distribution functions, we take the 
mass difference
between the heavy meson and its heavy constituent(s) 
as shape parameter. These distribution
functions follow the consensus that the smaller the mass difference 
the sharper 
the distribution functions. Using these distribution functions, we obtain
\begin{equation}  
{\rm Br}({B}^0 \to J/\Psi D^* )=8.50\times 10^{-8},~~~
P_{L}=\frac{\Gamma_{L}}{\Gamma}=0.395.
\end{equation}
 
Since ${B}^0$ decays with three charm quarks in its final states, it 
could be taken as a probe of strong interactions, especially  hadron 
dynamics.
We extend our calculations to 
${B}^0 \to J/\Psi D, \, \eta_c D$ and $\eta_c D^*$ 
decays. The amplitudes for these decays can  be obtained 
through the following replacements
in Eqs. 3 to 7
\begin{eqnarray}
{\bf  \Psi}_{D^*} \longrightarrow {\bf \Psi}_{D}
&=& \frac{i}{4N_c}\gamma_5 (\spur{P_{D}} + M_{D}) \phi_{D}(x) f_{D}, 
\nonumber  \\
{\bf  \Psi}_{\Psi} \longrightarrow  {\bf  \Psi}_{\eta_c}
&=& \frac{i}{4N_c}\gamma_5 (\spur{P_{\eta_c}} + M_{\eta_c})\phi_{\eta_c}(x) 
f_{\eta_c}.
\end{eqnarray}   
Using $f_{D}=200~{\rm MeV}$, $f_{\eta_c}=335~{\rm MeV}$, the 
branching ratios 
are estimated  to be 
\begin{eqnarray}    
{\rm Br}({B}^{0}\to J/\Psi D)&=&7.28\times10^{-8}, \nonumber \\
{\rm Br}({B}^{0}\to \eta_c D^{*})&=&1.39\times10^{-7}, \nonumber \\  
{\rm Br}({B}^{0}\to \eta_c D)&=&1.52\times10^{-7}.  
\end{eqnarray}

In summary, we have studied the decays ${B}^{0}\to J/\Psi (\eta_c) D^{(*)}$ 
within the conventional theoretical framework. The branching ratios of these
decays
are estimated to be
around $10^{-7}\sim 10^{-8}$. ${B}^0 $ decays to  $J/\Psi D^{(*)}$ 
can not account for the excess for slow $J/\Psi$ 
as indicated by the CLEO measurement of the $J/\Psi$ momentum spectrum 
in $B$ inclusive 
decays. Experimentally, inclusive decays of $B$ mesons to charmonium 
could be well 
studied at BaBar and Belle, and it is important to confirm whether the slow
$J/\Psi$ hump exists with refined measurements. If the excess persists, 
it would be 
hard to explain the phenomena within the conventional theoretical framework 
for hadron dynamics. As shown in here, our numerical results are 
rather stable under the change of parameters.
$If$ these exclusive decays were observed to be abnormally large,
say, of order $10^{-4}\sim 10^{-5}$, it would challenge the conventional 
theoretical  
framework and bring forth new interesting QCD phenomena, like the
scenarios discussed in Ref. \cite{brod,hou} or the possibility raised
here, of the formation of a $\approx2$ GeV $s {\bar d}g$ 
hybrid state $H$ through
$B^0 \to J/\Psi H$.
Finally let us note, that multibody final states such as
$J/\Psi D^{(*)}+n\pi$, where $n_{\rm max}=1,~2$ for
$D^*,~D$, respectively, being on the edge of phase space,
are expected to be even smaller than
those with $n=0$.
\section*{Acknowledgments}

This work  is supported in part by the US-Israel Binational Science
Foundation and by the Israel Science Foundation.

\end{document}